# Antiferromagnetic ground state in NpCoGe

E. Colineau[1], J.-C. Griveau[1], R. Eloirdi[1], P. Gaczyński[1], S. Khmelevskyi[2], A. B. Shick[1,3] and R. Caciuffo[1]

[1] European Commission, Joint Research Centre, Institute for Transuranium Elements, Postfach 2340, 76125 Karlsruhe, Germany.
[2] Center for Computational Materials Science, IAP, Vienna University of Technology, Gusshausstrasse 25a, 1040, Vienna, Austria.
[3] Institute of Physics, Academy of Sciences of the Czech Republic, Na Slovance 2, 182 21 Prague, Czech Republic.

**Abstract**

NpCoGe, the neptunium analogue of the ferromagnetic superconductor UCoGe, has been investigated by dc-magnetization, ac-susceptibility, specific heat, electrical resistivity, Hall effect, $^{237}$Np Mössbauer spectroscopy and LSDA calculations. NpCoGe exhibits an antiferromagnetic ground state with a Néel temperature $T_N \approx 13$ K and an average ordered magnetic moment $<\mu_{Np}> = 0.80$ $\mu_B$. The magnetic phase diagram has been determined and shows that the antiferromagnetic structure is destroyed by the application of a magnetic field ($\approx 3$ T). The value of the isomer shift suggests a $Np^{3+}$ charge state (configuration $5f^4$). A high Sommerfeld coefficient value for NpCoGe (170 mJ mol$^{-1}$K$^{-2}$) is inferred from specific heat. LSDA calculations indicate strong magnetic anisotropy and easy magnetization along the c-axis. Mössbauer data and calculated exchange interactions support the possible occurrence of an elliptical spin spiral structure in NpCoGe.

The comparison with NpRhGe and uranium analogues suggests the leading role of *5f-d* hybridization, the rather delocalized character of *5f* electrons in NpCoGe and the possible proximity of NpRuGe or NpFeGe to a magnetic quantum critical point.

# I. INTRODUCTION

The discovery that ferromagnetic order and superconductivity coexist at ambient pressure in URhGe[1] and UCoGe[2] moved forward the frontiers in solid state physics. Indeed, such a coexistence is not allowed in the spin-singlet pairing s-wave superconductors described by the standard BCS theory[3], but was predicted to occur in unconventional spin-triplet pairing p-wave superconductors[4]. This striking phenomenon was first observed experimentally in $UGe_2$ under high pressure[5] and soon afterwards found at ambient pressure in URhGe whose orthorhombic TiNiSi-type structure presents U zig-zag chains like in $UGe_2$. URhGe is a weak ferromagnet ($T_C$ = 9.5 K, $M_0$ = 0.4 $\mu_B$) where superconductivity occurs at very low temperature ($T_{SC}$ = 0.26 K). Remarkably, the isostructural compound UCoGe also exhibits (very) weak ferromagnetism ($T_C$ = 3 K, $M_0$ = 0.05 $\mu_B$) and superconductivity ($T_{SC}$ = 0.7 K) at ambient pressure. Advances are largely due to the discovery of new systems, presenting similarities and discrepancies that add pieces to the puzzle, and possibly more easily accessible (ambient pressure, higher critical temperatures…) to a wide range of experimental techniques[6,7]. Other actinides like neptunium and plutonium also form intermetallic compounds with highly interesting physical properties, like multipolar ordering (magnetic triakontadipoles) in $NpO_2$ [8,9] or superconductivity with critical temperatures one order of magnitude higher than in uranium heavy fermions superconductors ($PuCoGa_5$, $T_{SC}$ = 18.5 K [10]; $PuRhGa_5$, $T_{sc}$ = 8.7 K [11], $NpPd_5Al_2$, $T_{sc}$ = 4.9 K [12] and recently $PuCoIn_5$, $T_{sc}$ = 2.5 K [13] and $PuRhIn_5$, $T_{sc}$ ~ 1-2 K [14]). However, transuranium compounds are much less documented than uranium systems, due to the difficulty to handle them, their limited availability and their access restricted to a few licensed laboratories. In the TiNiSi-type structure, only NpNiSn was previously reported in the literature, as an antiferromagnet with $T_N$ = 37 K [15]. In this context, we have undertaken the investigation of the neptunium analogues of URhGe and UCoGe. NpRhGe orders antiferromagnetically below $T_N$ ≈ 21 K with an ordered moment $\mu_{Np}$ = 1.14 $\mu_B$ and no hint of superconductivity has been found down to 1.8 K. The specific heat indicates a high Sommerfeld coefficient value $\gamma$ = 195 mJ mol$^{-1}$K$^{-2}$ [16]. Preliminary measurements[17] have shown that NpCoGe is an antiferromagnet ($T_N$ ≈ 13 K). In the present paper, we report on the extensive study of the magnetic and electronic properties of NpCoGe, as inferred from dc and ac magnetization, specific heat, electrical resistivity, $^{237}$Np Mössbauer spectroscopy and LSDA calculations.

## II. EXPERIMENTAL

A polycrystalline NpCoGe sample was prepared by arc melting the constituent elements (Np: 99.97%, Co: 99.998% and Ge: 99.999%) in stoichiometric amounts under a high purity argon atmosphere (6N) on a water cooled copper hearth using a Zr getter. The as-cast sample wrapped in a Ta foil was annealed for one month at 850°C in a vacuum sealed quartz tube. Powder diffraction patterns recorded on a Bragg-Brentano Siemens diffractometer installed inside a glove-box confirmed the TiNiSi-type structure (space group Pnma, N. 62). The diffraction patterns were analysed by a Rietveld-type profile refinement method using the program Topas[18] version 4.1 ($R_{wp}$=3.95%, Gof=1.80). The resulting room temperature lattice parameters of the orthorhombic cell are a = 6.8837(2) Å, b = 4.2307(8) Å and c = 7.2144 (6) Å and the atomic positions are listed in Table I.

Table I : Atomic positions in NpCoGe

| Atom | x | y | z | Occupancy |
|---|---|---|---|---|
| Np | 0.00826 | 0.25000 | 0.70369 | 1 |
| Co | 0.29610 | 0.25000 | 0.39026 | 1 |
| Ge | 0.19752 | 0.25000 | 0.07568 | 1 |

Dc-magnetization and ac-magnetic susceptibility measurements were carried out on a 48.6 mg sample using a MPMS-7 SQUID magnetometer and a PPMS-14 device, respectively. The specific heat experiments were performed using the relaxation method on a 4.6 mg sample in a PPMS-9 system within the temperature range 1.8 – 300 K and in magnetic field up to 9 T. Electrical resistivity, magnetoresistivity and Hall effect were measured in the temperature range 1.8-300 K and in magnetic fields up to 14 T using a PPMS-14 setup and performed by a five dc probe technique voltage measurement. Two annealed samples with typical size 1.5×0.5×0.3 mm$^3$ have been polished with parallel faces to better determine the form factor. Electrical contacts have been ensured by using silver epoxy (Dupont 4929), at the contact between silver wires of 100 μm and the samples' surface.

For electrical resistivity ρ and magnetoresistivity Δρ/ρ, the measured voltage V was determined in parallel with the applied current **J**, is in the polished plane. For longitudinal magnetoresistivity, the applied magnetic field **B** was parallel to **J** while for transverse magnetoresistivity, **B** was perpendicular to **J**. In the Hall configuration, V was determined perpendicularly to **J** and in the plane of the polished sample while **B** is applied perpendicularly to the plane of the sample. For all measurements, J = 5 mA.

The $^{237}$Np Mössbauer measurements were performed in zero-field only, on a powder absorber with a thickness of 120 mg of Np cm$^{-2}$. The Mössbauer source of 108 mCi of $^{241}$Am metal was kept at 4.2 K while the temperature of the absorber was varied from 1.7 to 25 K. The spectra were recorded with a sinusoidal drive system using conventional methods. The velocity scale was calibrated with reference to a NpAl$_2$ absorber ($B_{hf}$ = 330 T at 4.2 K).

### III. Results and discussion

*A. Magnetic measurements*

The temperature dependence of the magnetization (Fig. 1) shows that NpCoGe orders antiferromagnetically at $T_N$ = 13 K. The ac magnetic susceptibility (insert Fig. 1) does not show any frequency dependence in the studied range (18-9887 Hz) and is similar to the dc susceptibility. In the paramagnetic state, it can be fitted by a modified Curie-Weiss law (including a Pauli paramagnetism term $\chi_0$ = 9.23×10$^{-4}$ emu/mol) yielding an effective moment $\mu_{eff}$ = 2.2 $\mu_B$ and a paramagnetic Curie temperature $\Theta_p$ = -5.5 K. The effective moment, reduced compared to the free ion ($\mu_{eff}$ ≈ 2.7 $\mu_B$), is similar to that found in NpRhGe [16]. The negative interaction temperature confirms the presence of antiferromagnetic correlations in NpCoGe and its small absolute value compared to the Néel temperature is consistent with the fragile antiferromagnetic structure vanishing in applied magnetic fields.

The application of high magnetic fields destroys this transition and induces a ferromagnetic-like behaviour. The overlap of zero-field cooled and field cooled curves indicates the absence of significant domains effects. The magnetization as a function of magnetic field (Fig. 2) shows a clear deviation - above B ≈ 3 T - from the linear behaviour, interpreted as a metamagnetic transition and in agreement with the disappearance of the antiferromagnetic phase mentioned above. The magnetization reaches 0.58 $\mu_B$ at 14 T, but does not saturate. This can be due to magnetic anisotropy or to the occurrence of a complex magnetic structure with a ferromagnetic component.

*B. Specific heat measurements*

Fig. 3 shows the specific heat of NpCoGe measured from 300 K down to 1.85 K. The room temperature value amounts to C ~ 80 J mol$^{-1}$K$^{-1}$, slightly above the Dulong-Petit limit. At low temperature, an anomaly is observed at $T_N$ = 13 K, in agreement with the magnetization data. The application of magnetic fields (inset of Fig. 3) has three effects: 1) It slightly decreases the ordering temperature, which is typical for an antiferromagnetic transition, 2) the peak is broadened and collapsed, which is typical for a ferromagnet and 3) a second anomaly emerges for intermediate fields and vanishes for magnetic fields greater than 3.5 T. These observations confirm and complete the magnetic phase diagram inferred from magnetization results: NpCoGe orders at $T_N$ = 13 K, but the application of magnetic field destroys this antiferromagnetic phase (AF) and induces a ferromagnetic-like phase (F) (Fig. 4). However, the fact that a peak is still visible at the highest magnetic field (9 T) and shifted towards lower temperatures, suggests that the F-phase may not consist of a simple ferromagnetic arrangement of the magnetic moments, but rather of a ferrimagnetic or canted structure with a ferromagnetic component.

In the low-temperature region (below $T_N$), the zero-field specific heat (Fig. 5) can be accounted for by the sum of electronic, phonon, magnetic and nuclear contributions :

$$C = C_{el} + C_{ph} + C_{mag} + C_{nuc} \quad (1)$$

At such temperatures, the phonon contribution can be approximated by :

$$C_{ph} = \beta T^3 \quad (2)$$

The magnetic contribution of an antiferromagnet with linear magnon dispersion relation can be expressed by [19] :

$$C_{mag} = \alpha T^3 \quad (3)$$

The use of these formulae for the phonon and magnon contributions is validated by the experimental $T^2$ dependence of C/T. However, these two contributions cannot be separately extracted and only the sum (β+α) can be determined.

The electronic contribution is related to the Sommerfeld coefficient γ :

$$C_{el} = \gamma T \qquad (4)$$

From the fit, we infer γ = 170 mJ mol$^{-1}$K$^{-2}$.

Finally, the small increase of the specific heat at the lowest temperatures can be reproduced by a nuclear contribution due to the splitting of the I = 5/2 nuclear ground level of the $^{237}$Np nuclei by the hyperfine field :

$$C_{nuc} = C_2 \, T^{-2} \qquad (5)$$

It should be noticed that this latter contribution was not fitted to the curve but calculated using the average hyperfine field inferred from Mössbauer spectroscopy (see section D).

### C. Transport properties

#### 1. Electrical resistivity measurements

The electrical resistivity of NpCoGe is shown in Fig. 6. The room temperature absolute value amounts to 290 μΩcm, which is close to the values reported for the annealed UCoGe polycristals (300 μΩcm) [20,21] and much larger than for NpRhGe (~100 μΩcm) [16]. From 300 K (RT) down to $T_N$, the resistivity decreases monotonously but the compound presents a very small Residual Resistivity Ratio (RRR ~1.2) even at the lowest temperature achieved (~1.8 K) and despite the annealing thermal treatment. The magnetic transition is marked by the appearance of a small peak with an upturn around $T_N$~13 K. Usually, thermal treatments on crystals improve transport properties quality by reducing defects and disorder in the material[22] but it has also been reported that very long annealing treatments on UCoGe can decrease the quality of the transport properties by a partial loss of Ge [22-24]. On the other hand, these thermal treatments should reinforce potential superconductivity or magnetic signatures[23] and the Residual Resistivity Ratio (RRR ~30 are commonly reached[22,23]). However, in NpCoGe, the observed RRR remains small and magnetic order features are unchanged before and after the annealing process. The shape of the resistivity curve in NpCoGe is similar to those observed in "high resistivity" – even after annealing - samples of UCoGe [24,25].

The magnetic contribution to the electrical resistivity, $\rho_{mag}$, can be determined assuming that the phonon scattering part to the resistivity of NpCoGe, $\rho_{ph}$, is the same as in a non-magnetic reference. ThCoGe [26] crystallizes into a different structure (AlB$_2$-type) but represents the closest non-magnetic analogue of uranium and neptunium counterparts. The inset of Fig. 6 shows the obtained $\rho_{mag}$ for NpCoGe, which displays a clear maximum around 40 K, as observed in UCoGe and also some other magnetically ordered neptunium-based compounds such as NpCoGa$_5$ [27] or NpNiSi$_2$ [28]. Above this temperature, $\rho_{mag}$ shows a negative logarithmic dependence characteristic of dense Kondo systems which may be well approximated by the formula $\rho = \rho_{HT} - c_K \ln T$ related to the scattering of conduction electrons by defects and imperfections in the crystal lattice on one side ($\rho_{HT}$) and disordered spins and Kondo impurities on the other side ($c_K$). Least-squares fit of this expression to the experimental data (inset of Fig. 6) yields the parameters $\rho_{HT}$ = 425 µΩcm and $c_K$ = 46.7 µΩ.cm which are clearly reduced compared to the values reported for UCoGe [24].

*2. Magnetoresistivity*

Fig. 7 shows the longitudinal resistivity of NpCoGe in magnetic fields up to 14 T. The upturn related to the magnetic transition vanishes above 4 T. The low-temperature (1.8 K) resistivity decreases dramatically with increasing fields and collapses down to ~ 50 % of its zero-field value at high fields. The inset of Fig.7 shows that the collapse of resistivity starts at the critical field 3.7 T (mid-point of the collapse ~ 5.5 T), in relative good agreement with magnetisation and heat capacity results. The slight difference is likely to arise mainly from the anisotropy of the orthorhombic structure – as observed in UCoGe by magnetization, critical field and thermal conductivity (M, H$_{c2}$, κ) [20,22,23]. At 50 K, we observe a slight negative magnetoresistivity term, close to -2 % at 14 T, with a B$^2$ dependence coming from the suppression of incoherent Kondo scattering in applied magnetic field[29,30].

Fig. 8 shows the transverse magnetoresistivity as a function of magnetic field up to 9 T. On isotherms from 2 K to 7 K, two peaks are observed, around 3.5 T and 5.3 T, respectively. The transverse magnetoresistivity increases with temperature, like the longitudinal magnetoresistivity. However, from 8 K to 15 K, the transverse magnetoresistivity progressively decreases and the 5.3 T peak vanishes (or shifts to higher fields). These differences between transverse and longitudinal magnetoresistivity are probably due to the magnetocrystalline anisotropy of this orthorhombic structure.

*3. Hall effect:*

Fig. 9 shows the temperature dependence of the Hall coefficient ($R_H$) of NpCoGe in a field of 9 T. At 250 K, $R_H$ is positive and amounts to $\sim 10^{-9}$ m$^3$C$^{-1}$. Its value continuously increases with decreasing temperature down to the Néel temperature, $T_N$=13 K presenting a huge maximum, reminiscent of the magnetic transition of the magnetic susceptibility $\chi$. It is interesting here to note that the sign of $R_H$ does not change in the whole temperature range meaning that the balance between the two types of carriers (holes and electrons) does not change drastically. The Hall coefficient includes two terms, $R_H = R_0 + R_S$, where $R_0$ is the ordinary Hall coefficient and $R_S$ is the extraordinary or anomalous Hall coefficient. According to the model of Fert and Levy[31] the temperature dependence of the Hall coefficient of a Kondo lattice above the Kondo temperature may be expressed as:

$R_H(T) = R_0 + \gamma_1 \rho_{mag}(T) \chi^*(T)$

where the first term, $R_0$, describes the Hall effect due to Lorenz motion of carriers and/or residual skew scattering by defects and impurities, while the second term comes from skew scattering by Kondo impurities. In this formula, $\chi^*(T)$ is the reduced susceptibility approximated by $\chi(T)/C$, where C is the Curie-Weiss constant and $\rho_{mag}(T)$ is the magnetic contribution to the electrical resistivity. The inset of Fig. 9 shows the scaling of the Hall effect $R_H$ versus $\rho_{mag}(T) \chi^*(T)$. Least-squares fitting of the above equation to the experimental data in the temperature range 50–300 K leads to the following values: $R_0$=-1.41 x 10$^{-9}$ m$^3$C$^{-1}$ and $\gamma_1$=0.154 K T$^{-1}$. The negative sign of $R_0$ indicates that the ordinary Hall effect is dominated by electron contributions. The Hall constant of UCoGe has not been reported in the literature, but some data point to a low density of carriers in this analogue compound also[24,32]. A simple one-band model provides an estimation of the concentration of free electrons close to $0.44\times10^{22}$ cm$^{-3}$, which should be considered as the upper limit of the actual carrier concentration in NpCoGe in the normal state and corresponds to ~0.57 free electron per formula unit. This half electron value at high temperature is a rough estimation of carriers participating to transport as the Hall coefficient is the result of a combination of electrons and holes' contribution with different carrier velocities and relaxation times. The $\gamma_1$ constant value is just twice larger than the one obtained for cerium-based heavy fermion compounds such as CeCu$_6$, CeAl$_3$ or UPt$_3$ [33] and of the same magnitude as CeRu$_2$Si$_2$ [34]. We may also note here that the extraordinary contribution clearly dominates R$_H$ in the non-ordered state, illustrating the predominance of magnetic contributions to the transport properties of NpCoGe.

## D. $^{237}$Np Mössbauer measurements

In the paramagnetic state (see Fig. 10, upper spectrum at T=25 K), $^{237}$Np Mössbauer spectra consist of a single site experiencing a pure quadrupolar splitting with a quadrupolar interaction parameter $|e^2qQ|$ = 18.0 mm s$^{-1}$ and an asymmetry parameter $\eta$ = 0.42. The lines are narrow (linewidth W=2.94 mm s$^{-1}$) and the isomer shift is found to be $\delta_{IS}$ = - 5.3 mm s$^{-1}$ with reference to the standard absorber NpAl$_2$. This value suggests the occurrence of Np$^{3+}$ ($^5$I$_4$) ions in NpCoGe with the electronic configuration $5f^4$.

In the magnetically ordered state (Fig. 10, lower spectrum at T=1.7 K), the Mössbauer spectrum reveals, as expected, a magnetic splitting resulting from the full degeneracy lift of nuclear levels (I=5/2) by the electronic hyperfine field (B$_{hf}$). Furthermore, the complex spectrum cannot be reproduced assuming a unique value of B$_{hf}$, but rather requires the combination of 4 sub-spectra with different values of B$_{hf}$ and same intensity ratio (Table II). This suggests the occurrence of a modulated magnetic structure with a maximum moment $m_{Np}^{max}$ = 1.0 $\mu_B$ and an average moment $<m_{Np}>$ 0.8 $\mu_B$ as illustrated in Fig. 11. The value of the quadrupolar interaction parameter in the magnetic state is about one third of its value above T$_N$, suggesting that the magnetic moments are oriented at 42° or 71° from V$_{zz}$, the principal axis of the electric field gradient (see for example ref. [35] for a thorough discussion). The direction of the V$_{zz}$ axis cannot be inferred from the low crystallographic symmetry of NpCoGe [15], but it is calculated (see next section) to point in the (a,c) plane at an angle of 13.5° from the c-axis. This would be compatible with the spin spiral model developed hereafter (moments in the (a,c) plane and forming 45° angles between them).

Table II. Mössbauer parameters obtained at 1.7 K and 25 K. The ordered magnetic moments are inferred from the hyperfine magnetic field via the Dunlap-Lander relationship.

| T(K) | $B_{hf}$ (T) | $m_{Np}$ ($\mu_B$) | Intensity (%) | $\delta_{IS}$ (mm s$^{-1}$) vs NpAl$_2$ | $e^2qQ$ (mm/s) | W(mm s$^{-1}$) | $\eta$ |
|---|---|---|---|---|---|---|---|
| 25 | 0 | 0 | 100 | -5.3 | 18.0$^a$ | 2.94 | 0.42 |
| 1.7 | 154 | 0.72 | 25 | -5.2 | -6.0 | 3.0$^b$ | 0.4$^b$ |
|  | 164 | 0.76 | 25 |  |  |  |  |
|  | 201 | 0.93 | 25 |  |  |  |  |
|  | 214 | 1.00 | 25 |  |  |  |  |

$^a$ Absolute value (the sign of $e^2qQ$ cannot be determined in the paramagnetic state)
$^b$ Values constrained in the fit

The four different values of the magnetic moments in the antiferromagnetic state of NpCoGe revealed by Mössbauer measurements (Table II) may also be accounted for by assuming an elliptical spin spiral ground state. Such a kind of magnetic structure usually appears due to competing exchange interactions, which stabilize the spiral magnetic configuration and strong spin orbit coupling/anisotropy on Np sites, which lead to the magnetic moment dependence on its orientation relatively to the easy magnetic direction. Examples of elliptical spin spiral magnetic structures are metallic Er [36], MnWO$_4$ [37], CdCr$_2$O$_4$ [38] and a few other materials.

The elliptical spin spiral structure **m(r)** can be expressed as:

**m(r)**=**p** sin(2π**q·r** + φ) + **t** cos(2π**q·r** + φ)

Here **r** is the vector defining the atomic position, **q** is the wave vector of the spiral, φ is the phase angle, and the semi-axes of the ellipse **p** and **t** are perpendicular to each other. In the simple case |**p**|=|**t**| the expression describes an ordinary spin spiral.

Since the Mössbauer spectra suggests four different values for the Np moment one can assume that the magnetic structure repeats itself every eight atoms in the direction of the spiral. The angle between the moments of neighboring Np atoms in the spiral is 45 degrees.

The best fit to the Mössbauer data given in Table II are provided by the values of the phase angle φ =16.04 degrees, |**p**| = 0.68 μ$_B$ and |**t**| = 1.01 μ$_B$. The resulting model magnetic structure is presented in Fig. 11, where the numbers enumerate consecutive Np atoms in the direction of the spiral. The moments Np(3,7) = 0.71 μ$_B$, Np(2,6) = 0.77 μ$_B$, Np(4,8) = 0.94 μ$_B$, and Np(1,5) = 0.99 μ$_B$ within 1% accuracy reproduce the values of Mössbauer data from Table II.

### E. Local spin-density approximation calculations

Electronic structure and magnetic properties of the itinerant magnet NpCoGe are investigated with the local spin density approximation and full-potential linearized augmented plane-wave method including spin-orbit coupling.

The Np-atoms in the unit cell form two "pairs" (1,2) and (3,4) (the atoms of the pair (1,2) will be further refered as Np-I, while (3,4) as Np-II) (see Fig. 12). The Np atoms within the pairs are connected by inversion I (1 → 2, 3 → 4), 180°-rotation $C_{2y}$ (1 → 2, 3 → 4) and mirror $\sigma_y$ (1-4 → 1-4), both accompanied by non-primitive translations. Other symmetry operations, 180°-rotations $C_{2x}$ (1 → 4, 2 → 3), $C_{2z}$ (1 → 3, 2 → 4), and mirrors $\sigma_x$ (1 → 3, 2 → 4), $\sigma_z$ (1 → 4, 2 →3) map the atoms Np-I of pair (1,2) to the atoms Np-II of the pair (3,4).

First, ferromagnetic order is considered for NpCoGe. The spin magnetization is fixed along each of the a[100]-, b[010]- and c[001]-axes. The crystal symmetry is reduced in the presence of spin-orbit coupling in order to preserve chosen components of the magnetization. The reduced symmetry {E,I} keeps the same full-potential expansion for all these directions of the spin magnetization. This allows to avoid systematic numerical errors when the difference in the total energies for different magnetization directions is calculated. Also, the reduced symmetry calculations allow for two pairs of Np-I atoms (1,2) and Np-II atoms (3,4) to become non-equivalent with respect to conventional space-group symmetry.

The conventional LSDA (von Barth-Hedin) band theoretical method is applied together with the relativistic full-potential linearized - augmented-plane-wave (FP-LAPW) method[39] to perform total energy electronic and magnetic structure calculations. The experimental crystal structure (both the lattice constants and internal parameters) were used in the calculations, and no structural optimization was performed.

Here, 320 special k-points in the irreducible 1/2 part of the Brillouin zone were used, with Gaussian smearing for k-points weighting. The "muffin-tin" radius values of $R_{MT}$ = 2.9 a.u. for Np, $R_{MT}$ = 2.15 a.u. for Co, $R_{MT}$ = 2.25 a.u. for Ge, and $R_{MT}(Ge) \times K_{max}$ = 7.875 (where, $K_{max}$ is the cut-off for LAPW basis set) were used.

The magnetic anisotropy energy (MAE) is calculated as the difference in the total energies for different orientations of the magnetic moment along a-, b- and c-axes, and is shown in Table III. Total energy calculations yield the c-axis to be an easy magnetization axis and a-axis as a hard magnetization axis. The energy difference between b- and c-axes is fairly small. It will be fair to call the b-c-plane as an easy-plane where the magnetization can be rotated.

The spin, orbital and total magnetic moments for Np and Co atoms together with the total magnetic moment (per f.u.) are shown in Table IV for three fixed directions of spin magnetization.

Total density of states (DOS) and spin-resolved densities of Np *f*-states and Co *d*-states are shown in Fig. 12, when the moment is along the easy c-axis.

The non-interacting value of Sommerfeld coefficient $\gamma$ = 84.9 mJ mol$^{-1}$K$^{-2}$ is evaluated from the value of DOS at the Fermi level. It is about a half of the experimental $\gamma$ = 170 mJ mol$^{-1}$K$^{-2}$ value and expected to increase due to the electron mass enhancement caused by dynamical electron interactions and electron-phonon coupling.

The electric field gradient (EFG, a second rank symmetric tensor of the second derivative of the Coulomb potential at the nucleus) can be obtained in the FP-LAPW calculations[40]. For the ferromagnetic case with the moments along the easy-axis, we calculate $V_{zz}$ = 6.253 x 10$^{21}$ V/m$^2$ and the asymmetry parameter $\eta$ = $[V_{xx} - V_{yy}]/V_{zz}$ of 0.54. The EFG principal z-axis is directed in a-c-plane, and rotated by 13.5° from the crystal c-axis.

TABLE III: Magnetic Anisotropy Energy (meV/f.u.) calculated from total energy differences for the magnetization directed along [100] (a-axis), [010] (b-axis), and [001] z (c-axis).

| ΔE | [100] – [001] | [010] - [001] | [100] – [010] |
|---|---|---|---|

| | | | | | | | | | |
|---|---|---|---|---|---|---|---|---|---|
| LSDA | | 9.47 | | | 1.31 | | | 8.15 | |

TABLE IV: Spin ($M_s$), Orbital ($M_l$) and Total ($M_j = M_s + M_l$) magnetic moments for Np atoms, and Total Magnetic Moment ($M^{Tot}$) per formula unit ($\mu_B$) calculated for three fixed directions of spin magnetization: x (a-axis), y (b-axis), z (c-axis).

| | \multicolumn{9}{c}{$M_s$ // a-axis [100]} |
|---|---|---|---|---|---|---|---|---|---|
| Moment | $M_s$ | | | $M_l$ | | | $M_J$ | | |
| Axis | x (a) | y (b) | z (c) | x (a) | y (b) | z (c) | x (a) | y (b) | z (c) |
| Np-I | 2.922 | 0 | 0 | -2.690 | 0 | -0.151 | 0.232 | 0 | -0.151 |
| Np-II | 2.922 | 0 | 0 | -2.690 | 0 | 0.151 | 0.232 | 0 | 0.151 |
| Co-I | -0.509 | 0 | 0 | -0.024 | 0 | -0.006 | -0.533 | 0 | -0.006 |
| Co-II | -0.509 | 0 | 0 | -0.024 | 0 | 0.006 | -0.533 | 0 | 0.006 |
| $M^{Tot}$ | 2.564 | 0 | 0 | -2.714 | 0 | 0 | -0.150 | 0 | 0 |
| | \multicolumn{9}{c}{$M_s$ // b-axis [010]} |
| Moment | $M_s$ | | | $M_l$ | | | $M_J$ | | |
| Axis | x (a) | y (b) | z (c) | x (a) | y (b) | z (c) | x (a) | y (b) | z (c) |
| Np-I | 0 | 2.922 | 0 | 0 | -2.727 | 0 | 0 | 0.195 | 0 |
| Np-II | 0 | 2.9220 | 0 | 0 | -2.727 | 0 | 0 | 0.195 | 0 |
| Co-I | | -0.486 | 0 | 0 | -0.016 | 0 | 0 | -0.502 | 0 |
| Co-II | | -0.486 | 0 | 0 | -0.016 | 0 | 0 | -0.502 | 0 |
| $M^{Tot}$ | 0 | 2.597 | 0 | 0 | -2.747 | 0 | 0 | -0.154 | 0 |
| | \multicolumn{9}{c}{$M_s$ // c-axis [001]} |
| Moment | $M_s$ | | | $M_l$ | | | $M_J$ | | |
| Axis | x (a) | y (b) | z (c) | x (a) | y (b) | z (c) | x (a) | y (b) | z (c) |
| Np-I | 0 | 0 | 2.923 | -0.203 | 0 | -2.703 | -0.203 | 0 | 0.220 |
| Np-II | 0 | 0 | 2.923 | 0.203 | 0 | -2.703 | 0.203 | 0 | 0.220 |
| Co-I | 0 | 0 | -0.448 | -0.021 | 0 | -0.013 | -0.021 | 0 | -0.461 |
| Co-II | 0 | 0 | -0.448 | 0.021 | 0 | -0.013 | 0.021 | 0 | -0.461 |
| $M^{Tot}$ | 0 | 0 | 2.637 | 0 | 0 | -2.716 | 0 | 0 | -0.079 |

*F. Calculations of the magnetic exchange interactions.*

In order to investigate the possible complex antiferromagnetic structure of NpCoGe we have performed scalar-relativistic calculations of the inter-atomic exchange interactions using first-principles Green-Function based Lichtenstein formalism[41] as implemented in Korringa-Kohn-Rostocker (KKR) band-structure method within Atomic Sphere Approximation (ASA) [42,43]. The radii $R_{at}$ of the ASA spheres was chosen to be $R_{Np}/R_{Co}=1.48$ and $R_{Np}/R_{Ge}=1.11$ and similar LSDA exchange correlation potential (Barth-Hedin) as in our FLAPW calculations has been employed.

The Lichtenstein formalism allows to calculate the exchange constants $J_{ij}$ of a classical Heisenberg Hamiltonian

$$H = - \sum_{i,j \in \{Np\}} J_{ij} \vec{e}_i \cdot \vec{e}_j \qquad (1)$$

where the summation is over Np sites of the underlying crystal lattice and $\vec{e}_i$ are the unit directional vectors of the spin magnetic moment on $i^{th}$ lattice site. In earlier work it was shown that the Hamiltonian (1) with calculated $J_{ij}$ according to the Lichtenstein scheme, predicts well the energy of spin spiral states in transition metals like Fe, Co and Cr [44] and non-collinear ground states in bcc Eu and (Mn,Cr)Au$_2$ alloys[45].

The application of this method to the case of NpCoGe requires however some justification since, in particular, Np has a large orbital moment, which is omitted in scalar relativistic KKR calculations. This justification is partly provided by the fact that localized Np moments indirectly interact with each other through polarization of the conduction band since in NpCoGe the Np-Np distances are relatively large and no direct overlap of *5f* shells is expected (see Table VI). This mediated interaction depends only on the value of the local spin moments and occurs via exchange of local spin states with conduction band, so-called RKKY type (see discussion in Ref. [46]). This fact is reflected in the large spin-polarization of the Co atoms in the ferromagnetic state of NpCoGe as can be seen from the results of relativistic FLAPW calculation (see Table IV). However, these large Co moments are non-intrinsic. They are induced in the ferromagnetic state by aligned Np moments and disappear in the paramagnetic state. The latter can be seen with conventional Disordered Local Moment (DLM) approach[47], which models the paramagnetic state above the magnetic ordering temperature. For the ferromagnetic state the scalar relativistic KKR-ASA yields 2.42 $\mu_B$/f.u. This value is very similar to the spin only magnetization of relativistic FLAPW calculations ~ 2.6 $\mu_B$/f.u (see Table IV). This total spin polarization arises from the Np spin moment – m(Np)=3.25 $\mu_B$, m(Co)=-0.72 $\mu_B$, and m(Ge)=-0.06 $\mu_B$. In the DLM state, where Np and Co moments are randomly oriented, the value of the Np spin moment remained almost unchanged: m(Np)$_{DLM}$=3.23 $\mu_B$, whereas the Co moment vanishes. The induced character of the Co moment in the ferromagnetic state is also verified by test calculations where Np moments were either randomly oriented or fixed to zero value whereas Co moments are allowed to be aligned with non-zero value. All these test calculations converge to the states with zero Co moment.

Thus the local character of Np moments and induced character of Co ones justify the use of Heisenberg Hamiltonian for mapping of the magnetic configuration energies with only Np moments included. The Hamiltonian (1) may therefore serve at least for crude quantitative analyses of the possible cause and kind of spiral state in NpCoGe discussed above.

We use the paramagnetic Disordered Local Moment state as a reference state for the calculation of exchange interactions constants between Np moments. The obtained values for the largest first nearest neighbor shells are presented in Table V.

Table V: Calculated Np-Np exchange interactions from Hamilatonian (1) for the first 9 nearest neighbor (NN) shells of Np atom in NpCoGe. The sign minus indicates antiferromagnetic character of the corresponding interaction. The third row gives the number of NN atoms in the respective shell.

| shell | 1NN | 2NN | 3NN | 4NN | 5NN | 6NN | 7NN | 8NN | 9NN |
|---|---|---|---|---|---|---|---|---|---|
| $J_{ij}$, mRy | 0.29 | 0.58 | -0.23 | -0.40 | 0.25 | -0.05 | 0.04 | -0.06 | -0.01 |
| atoms | 2 | 2 | 2 | 2 | 4 | 4 | 4 | 2 | 2 |

The first nearest neighbor 1NN and 8NN interactions are within the zigzag chain of Np atoms along the a-axis. The ferromagnetic 1NN and 2NN interactions dominate longer distance antiferromagnetic interactions in a-axis direction and lead to the ferromagnetic stacking of Np planes along the zigzag (a-axis) direction. The situation is completely different in bc-planes. In Fig. 13 we show the strongest interactions between zigzags Np moments. The zigzags of Np atoms are perpendicular to the plane of the figure. The 3NN and 4 NN antiferromagnetic interactions compete with ferromagnetic 2NN and 5NN interactions and it leads to the stabilization of the spin spiral configuration along the b-axis. We estimate the angle Θ of the spin spiral by considering the first 13 NN interactions (whose values are larger than computational accuracy of the method ~0.01 mRy) and minimizing the magnetic energy of the spiral along the b-direction:

$$E(\phi) = J_1 \cos(\phi) + J_2 \cos(2\phi) + J_3 \cos(3\phi)$$

where $J_1=2(J_{2NN}+J_{4NN}+2J_{5NN}+2J_{6NN})$, $J_2=2(J_{3NN}+J_{13NN}+2J_{7NN})$ and $J_3=2(J_{9NN}+J_{12NN})$ are the interplane couplings along the b-direction. The energy has a minimum at $\varphi = 38°$, which is close to the 45° angle of the elliptical spiral proposed to fit Mössbauer data.

Thus it appears that the calculated exchange interactions support the hypothesis of a spin spiral structure in NpCoGe and predicts its propagation direction to be along the b-crystallographic axis (moments in the (a,c) plane). On the other hand, relativistic calculations from previous section suggest that a-axis is the hard magnetization axis and c-axis is the easy one (Table III) and that the total magnetic moment of Np depends on its orientation with respect to the crystallographic axes (Table IV). Thus magnetic anisotropy may lead to the modulation of the magnetic moment amplitude in the spin-spiral configuration and explains its elliptic character.

The direction of the spin-spiral is defined by the exchange constants and not by the anisotropy. In any other possible direction of spin-spiral (different from b-axis) the strong ferromagnetic 1NN exchange interaction leads to the ferromagnetic minimum. The competition between anisotropy and exchange may define the polarization (polar angle of the spin spiral) but not its q-vector. The latter happens since for any value of the q-vector the contribution to the total energy from Magnetic Anisotropy Energy (MAE) is equal to the same quantity - half of the MAE value per magnetic site. Exceptions are the collinear cases q=1 (FM) and q=1/2 (AFM) with direction of the moments along an easy axis, where the MAE contribution is zero. Thus, for planar non-collinear spin-spiral to occur, one needs the energy of the spin-spiral state with respect to collinear configurations to be larger than the half value of the MAE energy. In our calculations the energy difference between the collinear antiferromagnetic state and spin-spiral states due to exchange is about 2.3 mRy/f.u. However, the difference between collinear FM state and spiral-state is just 0.035 Ry. If we compare this energy with the MAE from Table III (MAE for [100]-[001]) we find that exchange interactions alone does not override an anisotropy barrier 0.35 mRy/f.u to create a spin-spiral. Thus, although the calculated exchange interactions alone lead to the spin-spiral state in NpCoGe, the high energy of magnetic anisotropy calculated in previous section may prevent a spin spiral formation. The high value of MAE suggests that relativistic effects are rather strong in this system and may eventually signal for importance of pair-wise magnetic Moriya-Dzyaloshinskii interactions[48,49] - which we omit in our analyses. Due to low symmetry of the Np local atomic environment the conditions for onset of these interactions are satisfied for almost all Np-Np bonds (the absence of inversion symmetry with respect to the bond middle point). These interactions are able to produce a spin-spiral even in some low symmetry transition metal systems, like FeGe[50], where the relativistic effects are much weaker than in Np.

## IV. DISCUSSION

Mössbauer experiments suggest a complex magnetic structure for NpCoGe, with a modulated magnetic moment. Magnetization and specific heat measurements suggest that the metamagnetic phase (above ~3 T and up to our maximum experimental field, 14 T) is not a pure ferromagnetic structure. This situation is quite different than for NpRhGe where a unique value of the magnetic moment and robust antiferromagnetism in applied magnetic fields were observed. It can be mentioned that complex magnetic ordering (canted ferromagnetism) was initially reported in URhGe[51], but further studies concluded to collinear ferromagnetic order[52]. LSDA calculations have well reproduced the magnetic properties of URhGe[53] and UCoGe[32] and revealed strongly opposing spin and orbital parts and sizeable *5f-d* hybridization.
In the UTGe isostructural series, the tendency to magnetic ordering decreases as hybridization increases[54]. This is clearly illustrated by plotting the ordering temperature as a function of the shortest inter-actinide distance (Fig. 14). The weak ferromagnets URhGe and UCoGe are in the region where the magnetism collapses and where the superconducting dome appears.
Contrarily to their uranium counterparts, neither NpRhGe nor NpCoGe exhibit superconductivity (at least down to 1.8 K on standard quality polycrystals) but order magnetically at significantly higher temperatures than URhGe and UCoGe, respectively. As shown in Fig. 14, the distance of Np ions with their nearest neighbors is larger than the U-U distances in URhGe and UCoGe, hence the hybridization is expected to be weaker in the neptunium compounds, with more localized *5f* electrons and stronger magnetism.

Comparing NpRhGe and NpCoGe, we observe that the latter, with a shorter Np-Np distance, is a weaker magnet : the Néel temperature of NpCoGe is almost half of $T_N$ in NpRhGe (Fig. 13 and Table VI). Although the effective moments are similar in both compounds, the ordered moment is weaker in NpCoGe. Finally, the isomer shift is smaller in NpCoGe, suggesting the presence of more delocalized *5f* electrons (i.e. less screening of *s* and *p* electrons, increase of electronic density at the nucleus, decrease of $\delta_{IS}$).

Table VI. Ordering temperature, paramagnetic Curie temperature, ordered moment, effective moment, isomer shift, Sommerfeld coefficient and Np-Np distance in NpRhGe [15] and NpCoGe (this work).

| Compound | $T_{ord}$ (K) | $\theta_p$ (K) | $m_{ord}$ ($\mu_B$) | $m_{eff}$ ($\mu_B$) | $\delta_{IS}$ (mm s$^{-1}$) vs NpAl$_2$ | $\gamma$ (mJ mol K$^{-2}$) | $d_{Np-Np}$ (Å) |
|---|---|---|---|---|---|---|---|
| NpRhGe | 21 | -61 | 1.14 | 2.23 | -2.5 | 195 | 3.57 |
| NpCoGe | 13 | -5.5 | 0.8[1] | 2.2 | -5.3 | 170 | 3.51 |

[1] Average moment

Consistently, the paramagnetic Curie temperature is largely negative in the robust antiferromagnet NpRhGe but only slightly negative in the metamagnet NpCoGe. The Sommerfeld specific heat coefficient indicates a large density of states at the Fermi energy in both compounds, which is a common feature with URhGe and UCoGe.

It is also interesting to notice that, whereas URhGe and UCoGe are (weak) ferromagnets, UNiGe and UIrGe are antiferromagnets. In analogy, NpRhGe is a robust antiferromagnet, but NpCoGe is a weak antiferromagnet and field-induced ferromagnet.

From all these observations and trends described above, we can anticipate that NpRuGe should have a low ordering temperature, may be close to a magnetic quantum critical point and might be a good candidate for superconductivity. Efforts will be undertaken in this direction to further investigate the NpTGe system and unravel the intricate properties of the AnTGe isostructural family.

## V. SUMMARY

The experimental study and theoretical understanding of magnetic and superconducting ground states and their interplay is of major interest in condensed matter physics. The AnTGe (An = U, Np ; T = Pt, Pd, Ni, Ir, Rh, Co, Ru) isostructural series offers a unique opportunity to observe and investigate detailed properties and systematic trends in a system where ferromagnetism and superconductivity coexist at ambient pressure. Following our study of NpRhGe [15], we have investigated here NpCoGe and found that it orders antiferromagnetically at $T_N$=13 K with an average ordered moment $<m_{Np}>$ = 0.8 $\mu_B$. The weak antiferromagnetic interactions ($\theta_p$ = -5.5 K) are overcome by the application of a magnetic field ~3 T that induces a metamagnetic phase of, if not pure, dominant ferromagnetic character. The magnetic structure in zero-field is probably complex and intermediate between a pure sine and a squared modulation, with the possibility of spin spiral. NpCoGe appears as a more delocalized antiferromagnet than NpRhGe, which is consistent with the trend observed in UTGe analogues. This work allowed to complete the description of the magnetic and electronic properties of neptunium analogues of the ambient pressure ferromagnetic superconductors URhGe and UCoGe, draw a first picture of the trends in the NpTGe series and identify the next NpTGe system of interest to investigate, NpRuGe.

**Acknowledgements**

S.K. thanks the Austrian Science Fund (FWF) (SFB ViCoM F4109-N13). Stimulating discussions with Dr. J.-P. Sanchez and technical support from P. Amador Celdran, F. Kinnart, D. Bouëxière and G. Pagliosa are acknowledged. The high purity Np metals required for the fabrication of the compound were made available through a loan agreement between Lawrence Livermore National Laboratory and ITU, in the frame of a collaboration involving LLNL, Los Alamos National Laboratory and the US Department of Energy.

# Figures captions

Fig. 1: (Color online) Magnetization of NpCoGe versus temperature for various applied magnetic fields. Insert : ac magnetic susceptibility of NpCoGe versus temperature (logarithmic scale) for 18 Hz (blue circles) and 9887 Hz (black diamonds). The red (solid) line represents the Curie-Weiss fit in the paramagnetic state.

Fig. 2: (Color online). Magnetization of NpCoGe at 7 K as a function of magnetic field (up to 14 T, using the PPMS-14). Inset : Magnetization isotherms (2 K and 20 K) measured in the SQUID-MPMS-7.

Fig. 3: (Color online) Specific heat of NpCoGe in the 1.85-300 K temperature range. Inset : magnetic transitions for typical applied magnetic fields.

Fig. 4: (Color online) Magnetic phase diagram of NpCoGe as inferred from magnetization (squares) and specific heat (circles) measurements (see text).

Fig. 5: (Color online). Low-temperature specific heat of NpCoGe. The solid line represents a fit including the electronic, phonon, magnetic and nuclear contributions.

Fig. 6: (Color online). Electrical resistivity of NpCoGe. Inset: Magnetic resistivity of NpCoGe, $\rho_{mag}$, obtained by subtracting the phonon contribution represented by the resistivity of ThCoGe.

Fig. 7: (Color online). Longitudinal electrical resistivity of NpCoGe in different magnetic fields. Inset: Magnetoresistivity in the ordered (1.8 K) and non-ordered (50 K) state.

Fig. 8: Transverse magnetoresistivity from 2 K to 8 K and (inset) 8 K to 15 K.

Fig. 9: (Color online). Hall coefficient of NpCoGe determined at 9 T and molar magnetic susceptibility determined at 3 T. The inset shows the Hall coefficient $R_H$ scaled vs $\rho_{mag} \times \chi(T)^*$: a linear regime is visible from 50 K to room temperature.

Fig. 10:   $^{237}$Np Mössbauer spectra of NpCoGe recorded at 1.7 K and 25 K. .

Fig. 11:   Left : Possible magnetic modulation of the magnetic moment in NpCoGe as suggested by Mössbauer spectroscopy. The dotted line represents the average moment. Right : Magnetic elliptical spiral structure model for NpCoGe.

Fig. 12:  (Color online). Density of states (DOS) of NpCoGe (c-axis). The Np (f) contribution is larger (in both DOS and energy) than the Co (d) contribution. Insert : Schematic Crystal Structure of NpCoGe . The unit cell Np-atoms are marked as (1,2,3,4) and located at 4c-sites. The crystal axes (a,b,c) correspond to (x,y,z) axes.

Fig. 13:  The nearest neighbor shells of Np atoms in bc-plane. 2NN, 4NN and 5NN interactions contribute to 1st NN inter-plane coupling along the b-axis, whereas 3NN interactions contribute to 2nd NN inter-plane coupling. The AFM character of 3NN magnetic interactions is the reason for the instability of the FM configuration with respect to the formation of spin spiral along the b-axis.

Fig. 14: (Color online). Ordering temperatures of UTGe (T=Pt, Pd, Ni, Ir, Rh, Co, Ru) and NpTGe (T=Rh, Co) compounds versus the shortest inter-actinide (An-An) distance (data taken from [55]). The blue (light) area corresponds to magnetically ordered compounds, whereas the red (dark) area denotes the presence of superconductivity observed at low temperatures in URhGe and UCoGe.

Fig. 1

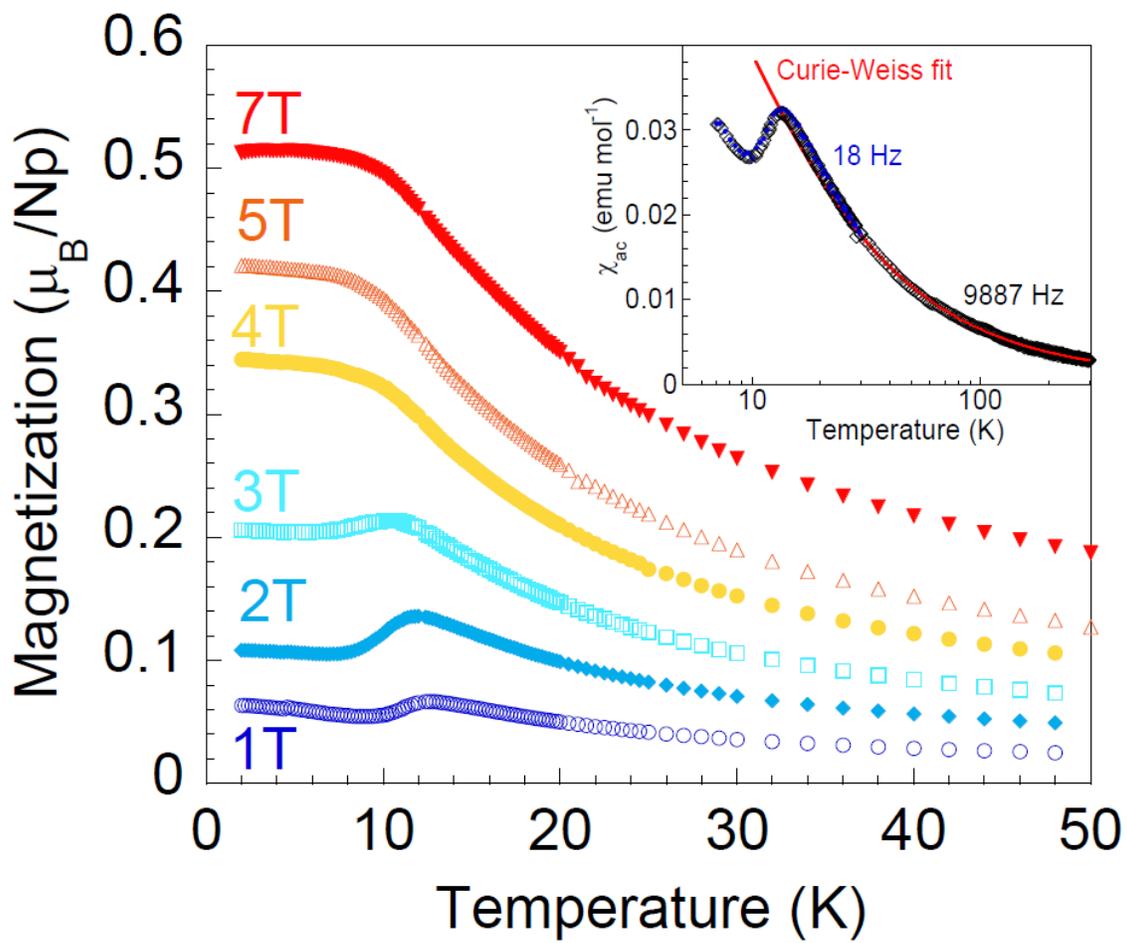

Fig. 2

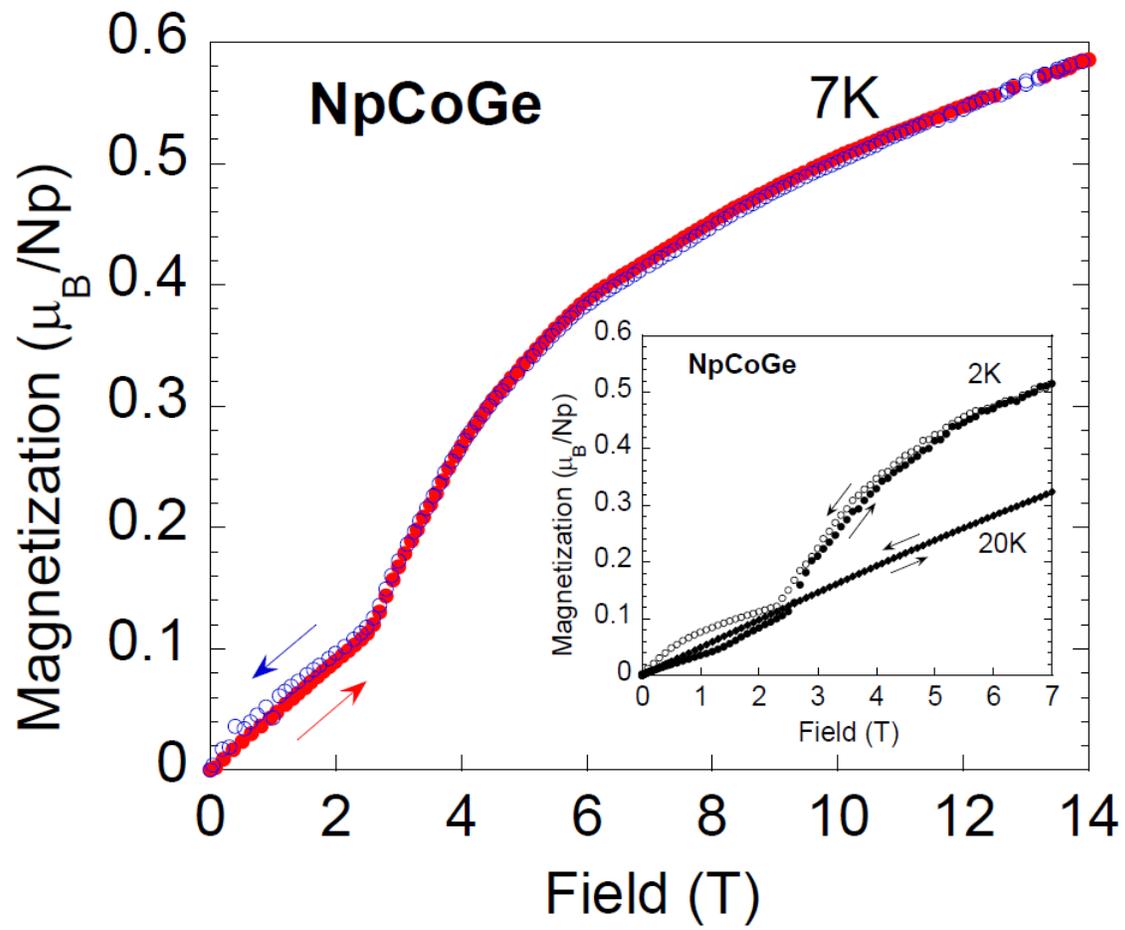

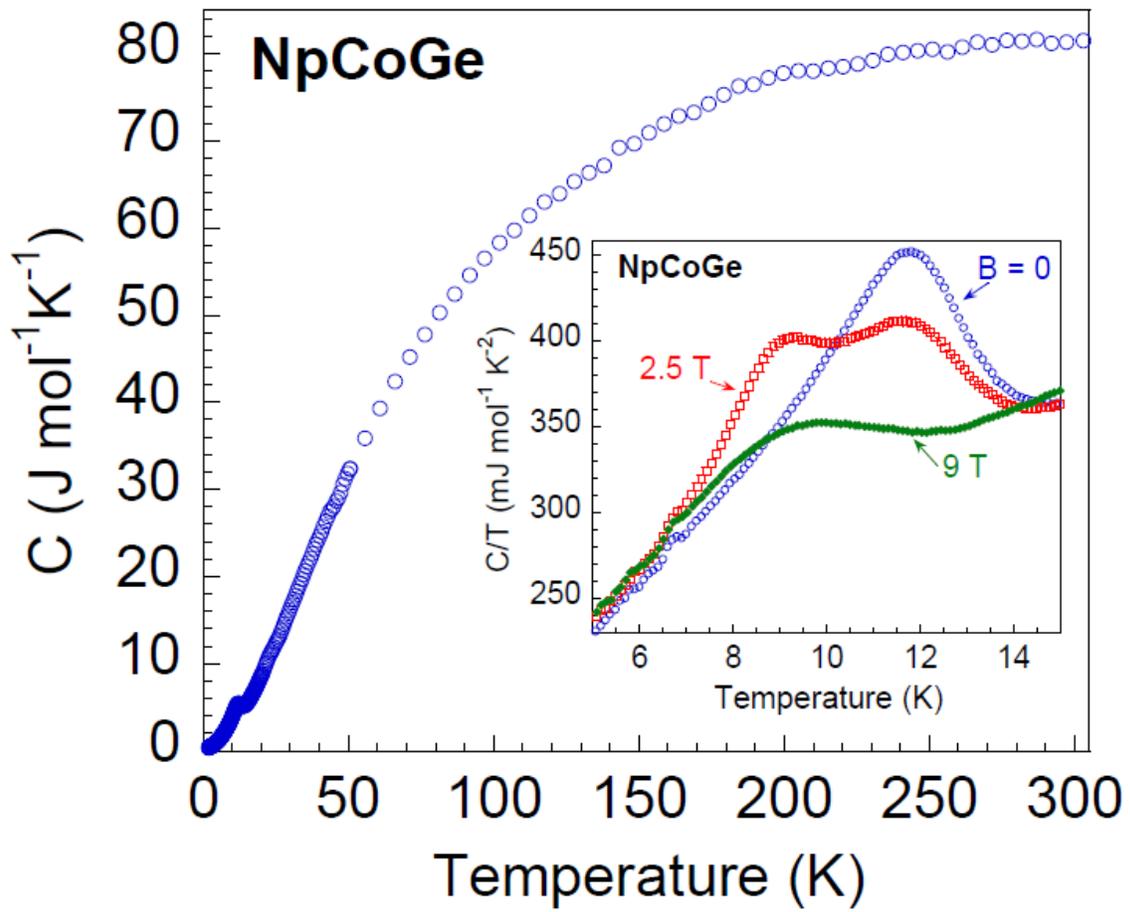

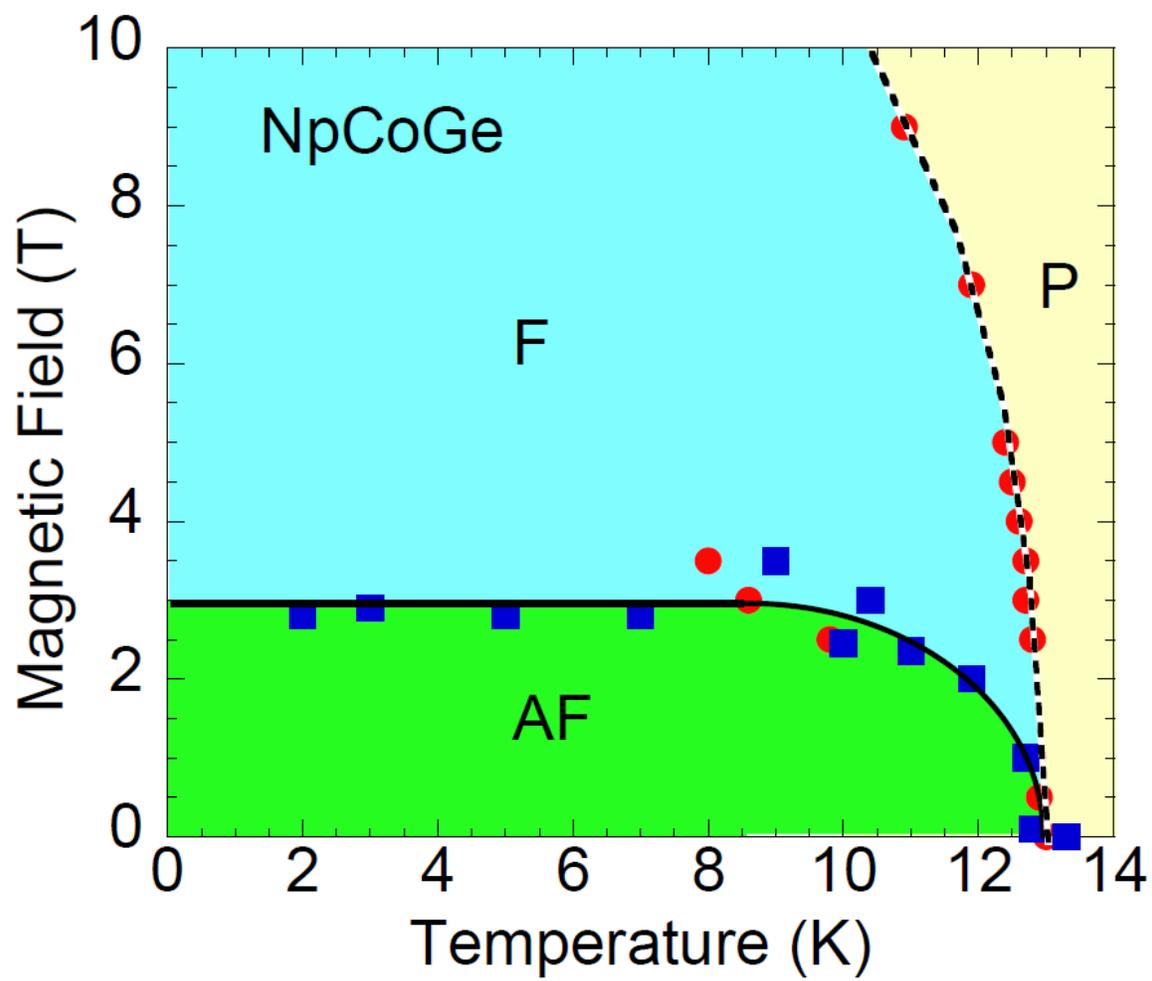

Fig. 4

Fig. 5

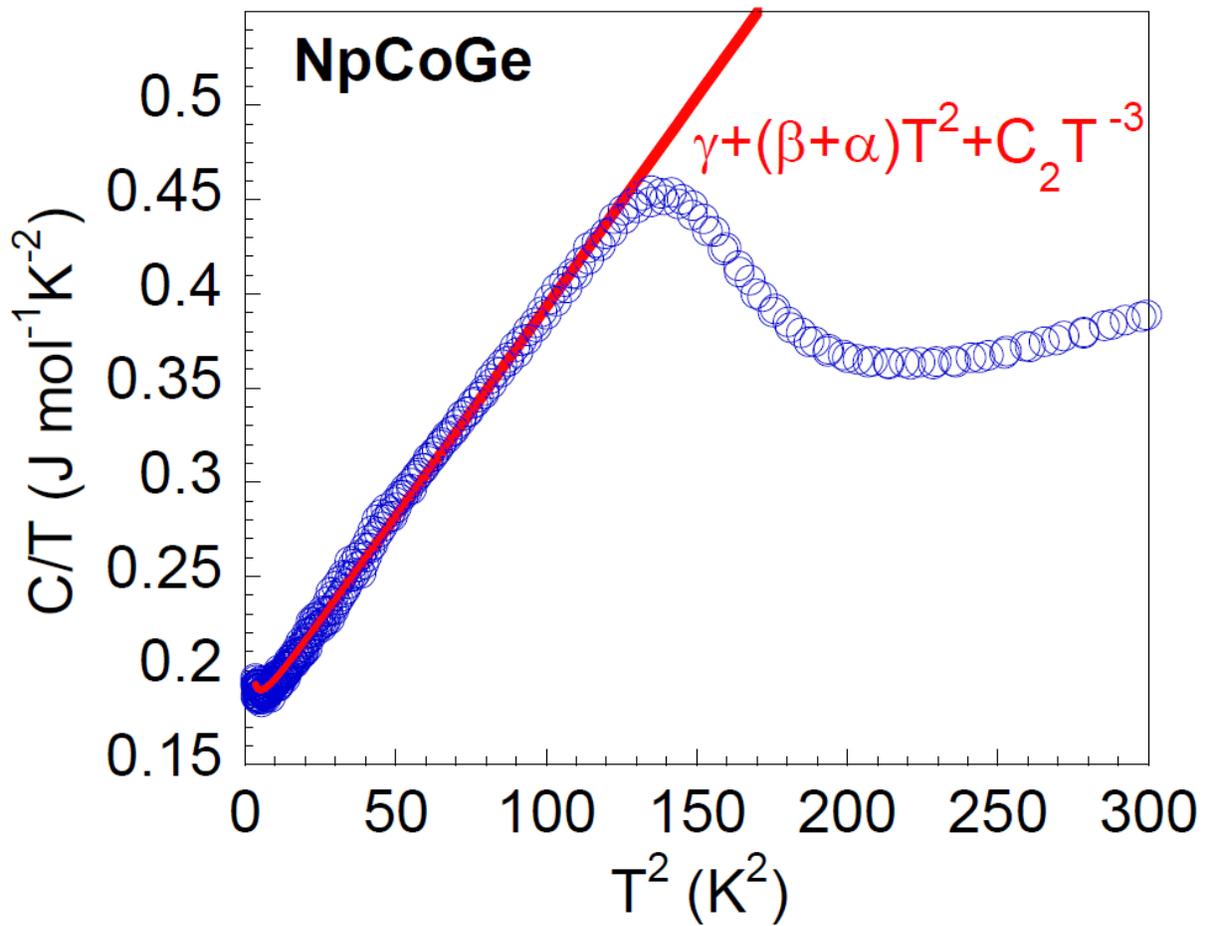

Fig. 6

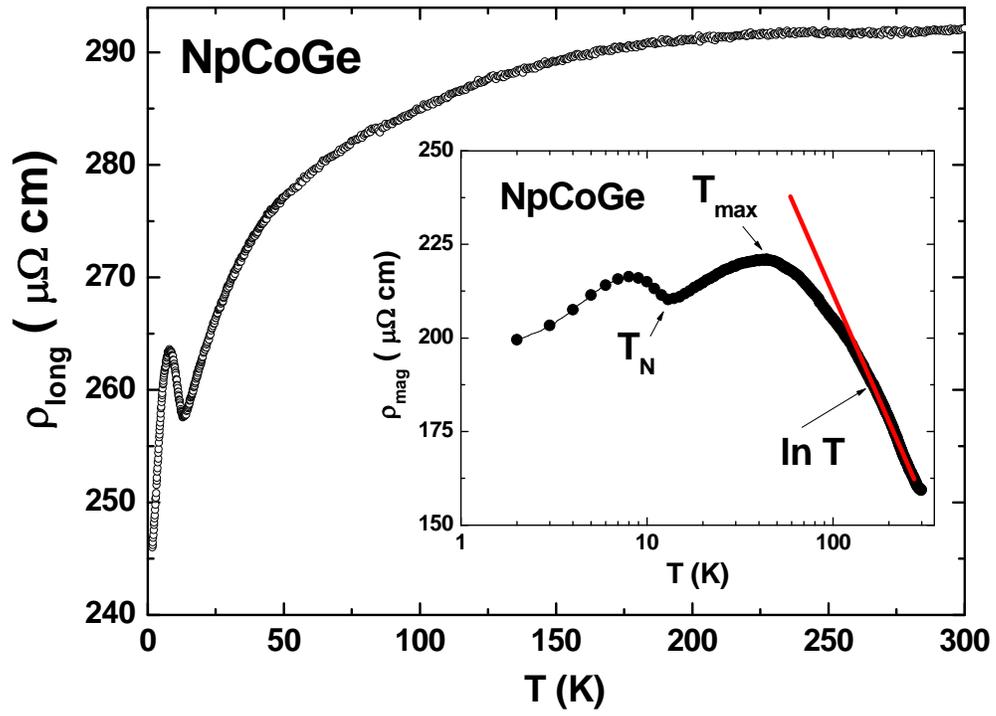

Fig. 7

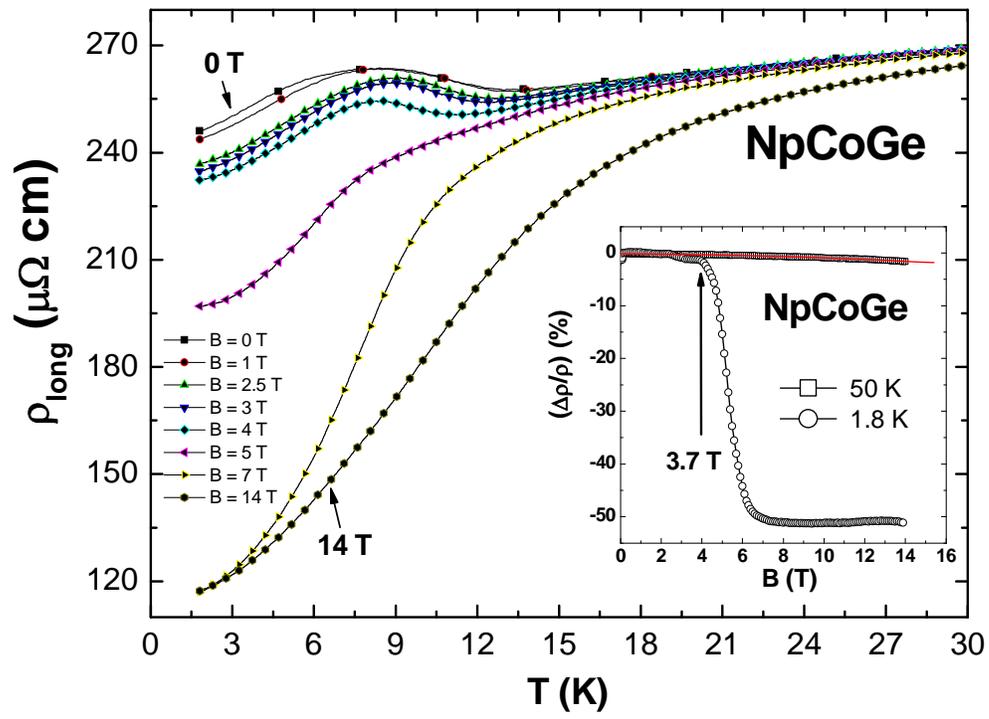

Fig. 8

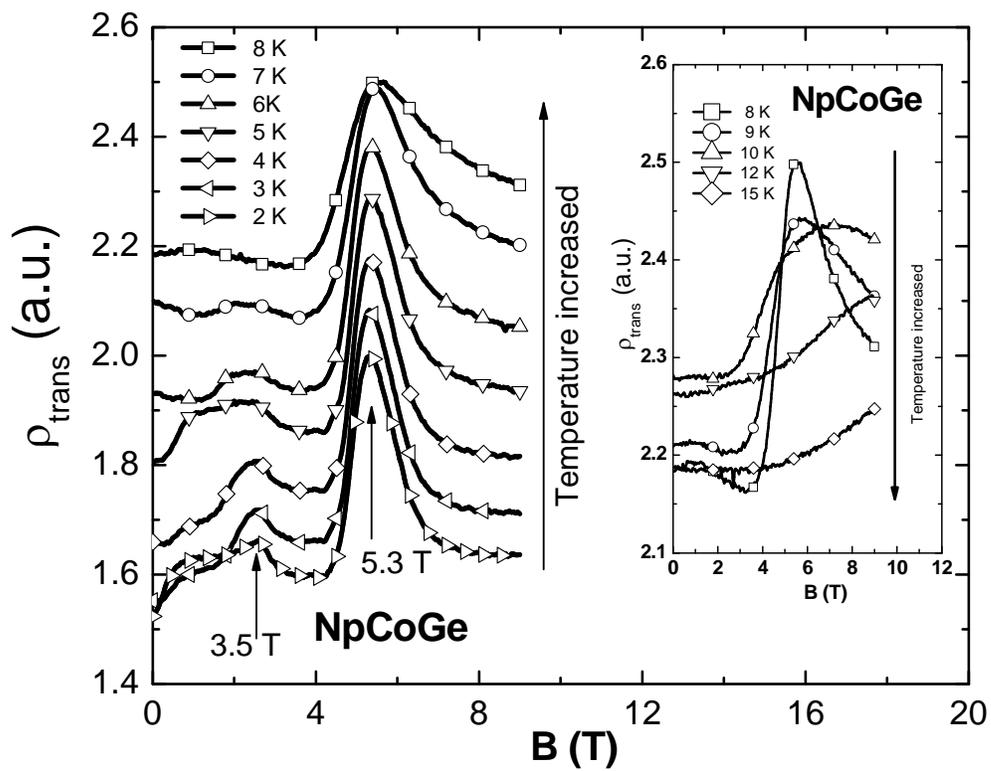

Fig. 9

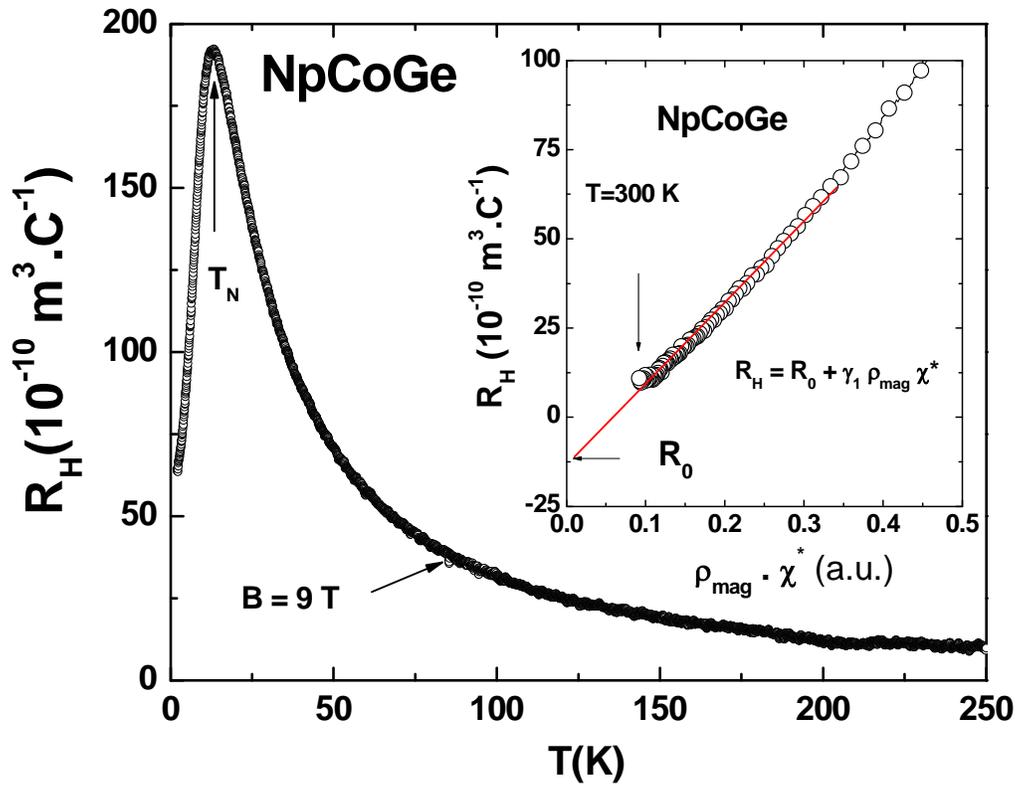

Fig. 10

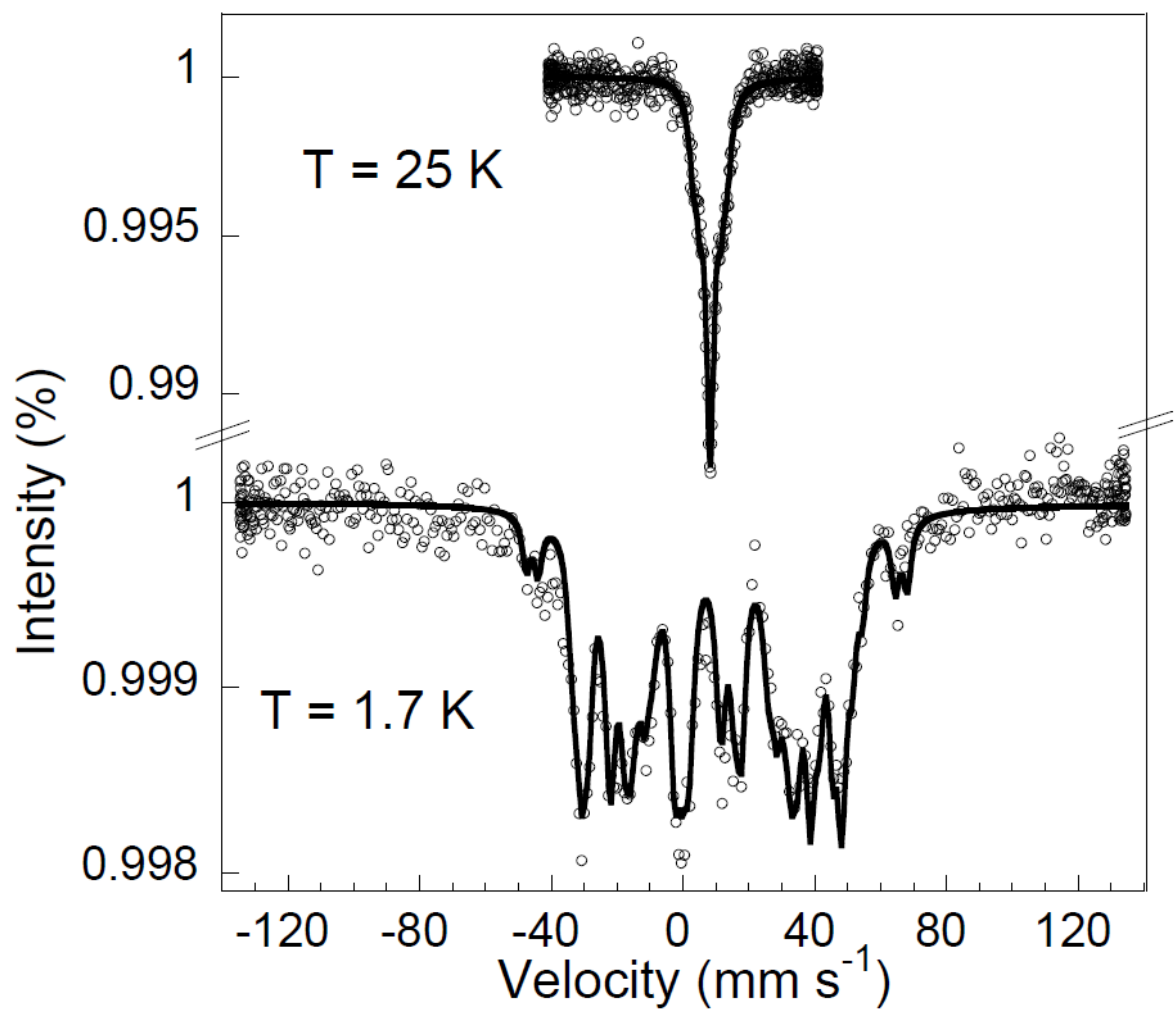

Fig. 11

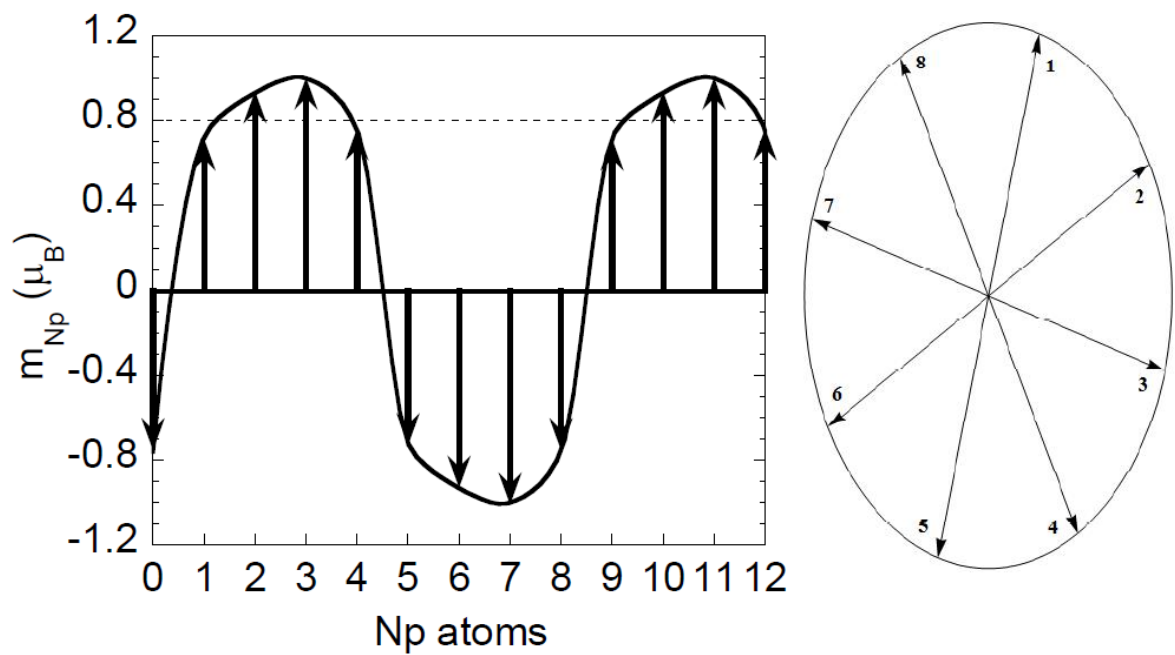

Fig. 12

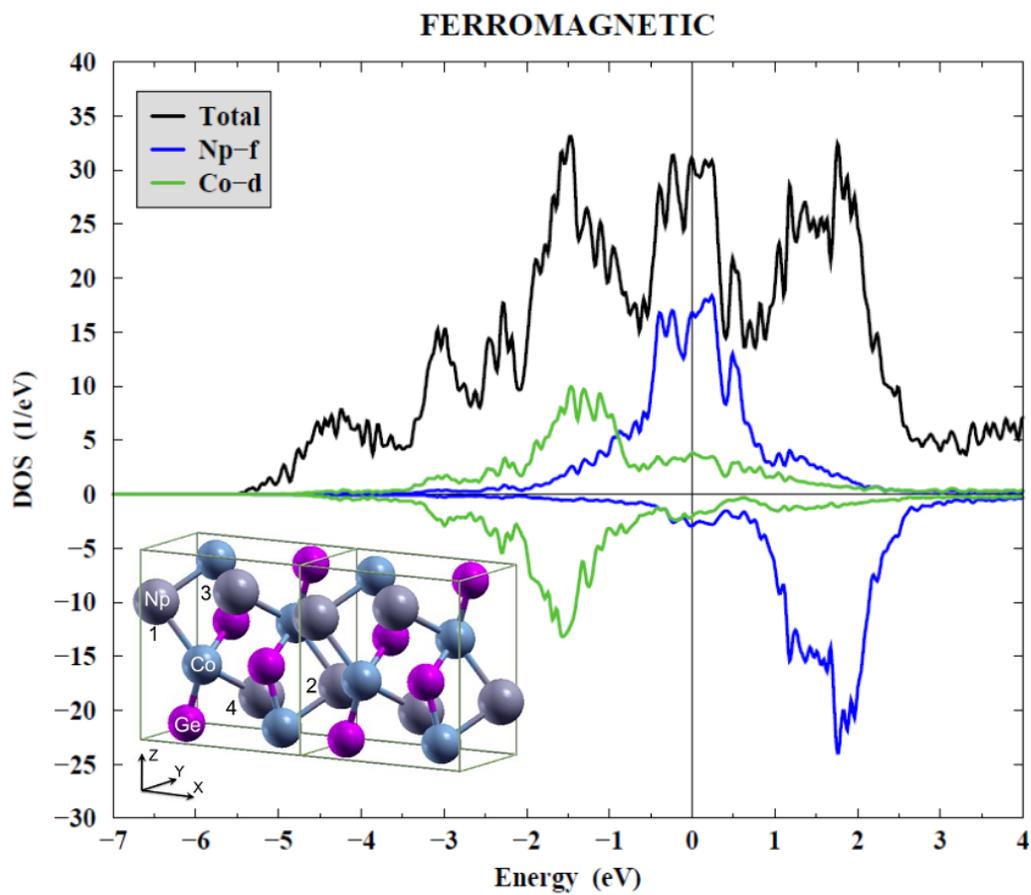

Fig. 13

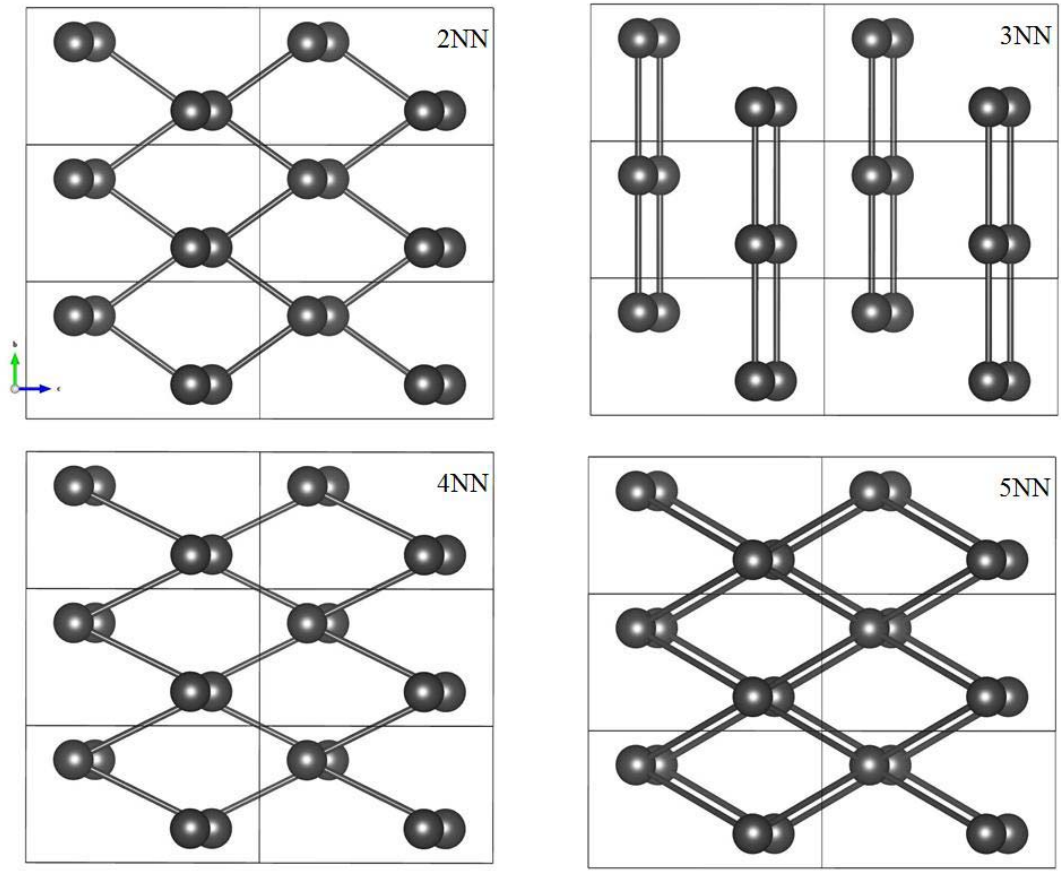

Fig. 14

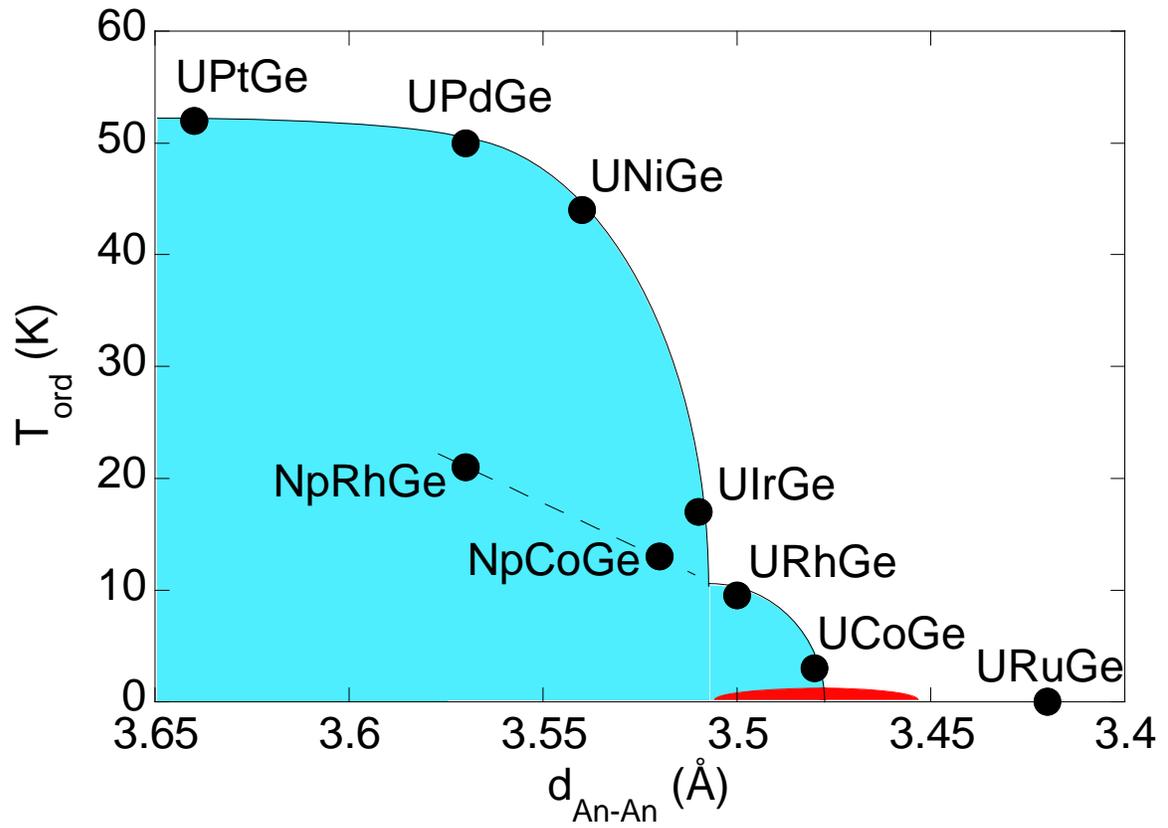